\documentstyle[preprint,aps]{revtex}
\begin{document}
\preprint{Draft : \today \ \ \ \ \ {\em [To be submitted to
Physical Review C]}}
 
\title{How to include a three-nucleon force into Faddeev equations for  
the 3N continuum: a new form}
 
\author{D. H\"uber$^1$, H. Kamada$^2$, H. Wita{\l}a$^3$, 
W. Gl\"ockle$^1$}

\address{$^1$Institut f\"{u}r Theoretische Physik II,
Ruhr-Universit\"{a}t Bochum, D-4630 Bochum, Germany}
\address{$^{2}$Paul Scherringer Institut (PSI), CH-5232 Villigen PSI,
  Switzerland} 
\address{$^3$Insitute of Physics, Jagellonian University,
Reymonta 4, PL-30059 Cracow, Poland}
\date{Received ---------1996}
\maketitle
 
\begin{abstract}
A new, more efficient  approach to include a three nucleon force into 
three-nucleon continuum 
calculations is presented. Results obtained in the new and our old approach
are compared both for elastic nucleon-deuteron scattering as well as for the 
breakup process. The advantages of the new scheme are discussed.
\end{abstract}
 \ \ \ \ \   \\
{\em PACS numbers : 21.60.Cs, 25.40.Kv, 27.30.+t}
\newpage
 
\vspace {.12in}
 
 
Three-nucleon forces (3NF) are getting more and more into the focus of 
few-nucleon studies, both theoretically\cite{ref1} and 
experimentally\cite{ref2}. Therefore optimal forms of equations should 
be used to handle those forces most efficiently.

One of the first 3N continuum calculations including a 3N force were 
performed in \cite{ref3}. Separable nucleon-nucleon (NN) forces 
together with a 
$\pi-\pi$ exchange 3NF\cite{ref4} averaged over spin and isospin 
degrees of freedom have been used. Thereby the standard Lovelace 
equations\cite{ref5} have been generalized to include a 3NF. Starting 
from the AGS form\cite{ref6} of the arrangement operators for 
identical particles and including a 3N force one gets\cite{ref7} 

\begin{eqnarray}
\label{e1}
U &=& PG_0^{-1} + (1+P)t_4 + PtG_0U + (1+P)t_4G_0tG_0U
\end{eqnarray}
Here $G_0$ is the free 3N propagator, t the NN t-matrix, $t_4$ the 
corresponding quantity now driven by the 3NF $V_4$ alone, and P the sum of a 
cyclical and anticyclical permutation of three 
 objects. That operator relation (\ref{e1}) 
has to be applied on a channel state $\vert \phi >$ corresponding 
to Nd scattering, 
which is a product of the deuteron wave function and the momentum eigenstate 
of the projectile. Once U has been determined one gets the operator for the 
breakup process by quadrature. For more details and our notation in general we refer to \cite{ref7}.

Eq.(\ref{e1}) has been solved in\cite{ref3} assuming NN t-matrices of finite 
rank. For the sake of a simple notation we just take the schematic form

\begin{eqnarray}
\label{e2}
t &=& \vert g > \tau <g \vert
\end{eqnarray}
which converts (\ref{e1}) into the structure of a single particle 
Lipmann-Schwinger equation

\begin{eqnarray}
\label{e3}
 X & = & ( Z^{(2)} + Z^{(3)} ) + ( Z^{(2)} + Z^{(3)} ) \tau X
\end{eqnarray}

where 

\begin{eqnarray}
\label{e4}
X & \equiv & < g \vert G_0 U G_0 \vert g > \\
\label{e5}
Z^{(2)} &=& < g \vert P G_0 \vert g >
\end{eqnarray}
and
\begin{eqnarray}
\label{e6}
Z^{(3)} &\equiv& {1\over{3}} < g \vert G_0 (1+P)t_4(1+P)G_0 \vert g >
\end{eqnarray}
In addition to the well known particle exchange term $Z^{(2)}$ there is now a potential 
term $Z^{(3)}$  driven by the 3NF. We used 
\begin{eqnarray}
\label{e7}
(1+P)t_4 &=& t_4(1+P) = {1\over{3}}(1+P)t_4(1+P)
\end{eqnarray}
In ref.\cite{ref3} Eq.(\ref{e3}) has been solved under 
the simplifying assumption $t_4=V_4$. 
We would like to mention another approach in the context of finite rank NN
forces \cite{ref3a}. There a simple ad hoc ansatz for $t_{4}$ has been
chosen. 

For general forces, which are not of finite rank, another set of 
equations within the Faddeev scheme was formulated\cite{ref7,ref8,ref9} and 
has been heavily used in \cite{ref10,ref11,ref12}. This is a 
set of two coupled 
equations in case of identical particles. They have the form

\begin{eqnarray}
\label{e8}
T &=& tP + tG_0T_4 + t P G_0 T \\
T_4 &=& (1+P)t_4 + (1+P) t_4 G_0 T \nonumber
\end{eqnarray}
Once T and $T_4$ are found the operator for the 3N breakup process is given 
as

\begin{eqnarray}
\label{e9}
U_0 &=& (1+P)T + T_4
\end{eqnarray}
and the operator for elastic Nd scattering as

\begin{eqnarray}
\label{e10}
U &=& PG_0^{-1} + PT + T_4
\end{eqnarray}
All these operator relations have to be applied onto the initial channel state 
$\vert \phi >$.

We have solved that coupled set\cite{ref10,ref11,ref12} by expanding $t_4$, 
T and $T_4$ in powers of $V_4$, which leads to a recursive set of 
equations\cite{ref8}. This is feasible and numerically precise but requires 
extensive resources in storage and computer time. 

For the 3N bound state in a momentum space treatment an algorithm has been 
proposed\cite{ref13}, which has been taken up again in \cite{ref14}. There 
the property of 3NF's applied  up to now have been used, that it can be split 
naturally into 3 parts, each one of which is symmetrical under exchange of 2 
particles, like the pair forces. For the $\pi-\pi$ exchange 3NF for instance 
that splitting occurs automatically, since there are three possibilities 
for choosing a nucleon which undergoes the (off-shell) $\pi-N$ scattering. 
Using that decomposition one can combine each of the three NN forces with 
a corresponding part of the 3NF having the same symmetry 
under particle exchanges. 
Then the derivation of the Faddeev equation from the Schr\"odinger equation 
follows exactly the same line as for NN forces alone and one arrives at one 
Faddeev equation using the identity of the particles. For the 3N bound state 
it has the form

\begin{eqnarray}
\label{e11}
\psi &=& G_0tP\psi + G_0(1+tG_0)V_4^{(1)}(1+P)\psi
\end{eqnarray}
This equation has been used in \cite{ref14} and later work, for instance 
in \cite{ref15}.

Based on that formal insight it is obvious that a corresponding form should 
exist for 3N scattering. The elaboration of that expectation is the content 
of the present work.

Let us split the 3NF into 3 parts

\begin{eqnarray}
\label{e12}
V_4 = \sum_{i=1}^3 V_4^{(i)}
\end{eqnarray}
as described above and let us arrange the 3N Schr\"odinger equation in 
integral form as 

\begin{eqnarray}
\label{e13}
\Psi &=& G_0 \sum_{i=1}^3 ( V_i + V_4^{(i)} ) \Psi
\end{eqnarray}
We used the usual ``odd man out'' notation for the NN forces. This is a 
correct homogeneous integral equation for a scattering state, which is 
initiated in the nucleon-deuteron (Nd) channel\cite{ref7}. 
Then splitting $\Psi$  into 
three Faddeev components ${\psi}_i$ one gets for the first component, for 
instance, 

\begin{eqnarray}
\label{e14}
\psi_1 &=& G_0 ( V_1 + V_4^{(1)} ) ( \psi_1 + \psi_2 + \psi_3 )
\end{eqnarray}
The NN t-matrix $t_1$ related to $V_1$ is then introduced in the normal manner 
and one ends up easily with 

\begin{eqnarray}
\label{e15}
\psi_1 &=& \phi_1 + G_0t_1(\psi_2 + \psi_3) + (1+G_0t_1)
G_0V_4^{(1)} ( \psi_1 + \psi_2 + \psi_3 )
\end{eqnarray}
Here $\phi_1$ is the initial channel state, as described above.

For identical particles

\begin{eqnarray}
\label{e16}
\psi_2 + \psi_3 &\equiv& P\psi_1
\end{eqnarray}
and dropping the index 1 we get

\begin{eqnarray}
\label{e17}
\psi &=& \phi + G_0tP\psi + (1+G_0t)G_0V_4^{(1)}(1+P)\psi
\end{eqnarray}
This is the Faddeev equation for 3N scattering initiated in a Nd channel.

Using $G_0t=GV$, where $G \equiv (E-H_0-V)^{-1}$, one can easily extract 
the asymptotic behaviour in the elastic and breakup channels\cite{ref16}. 
Thus the amplitude accompanying the familiar  outgoing wave in the 
hyperradius is

\begin{eqnarray}
\label{e18}
\tilde T &=& (1+tG_0)VP\psi + (1+tG_0)V_4^{(1)}(1+P)\psi \nonumber \\
~        &=& tP\psi + (1+tG_0)V_4^{(1)}(1+P)\psi
\end{eqnarray}

The full breakup amplitude constructed from all three Faddeev amplitudes is then 

\begin{eqnarray}
\label{e19}
 U_0 &=& (1 + P) \tilde T
\end{eqnarray}

Now we see directly comparing (\ref{e17}) and (\ref{e18}) that

\begin{eqnarray}
\label{e20}
\psi &=& \phi + G_0 \tilde T
\end{eqnarray}
and therefore

\begin{eqnarray}
\label{e21}
\tilde T &=& tP\phi + (1+tG_0)V_4^{(1)}(1+P)\phi + tPG_0\tilde T +
 (1+tG_0)V_4^{(1)}(1+P)G_0\tilde T
\end{eqnarray}
This is the searched for single equation, which replaces the coupled set 
(\ref{e8}). For $V_4^{(1)}=0$ it reduces to the form, which we always use 
in solving the 3N continuum for NN forces only\cite{ref11}.

The asymptotic behaviour of (\ref{e17}) in the elastic channel yields the 
operator for the elastic process

\begin{eqnarray}
\label{e22}
U &=& VP\psi + V_4^{(1)}(1+P)\psi
\end{eqnarray}

The elastic amplitude arises by projecting U from the left with the channel 
state $< \phi \vert$, which carries the correct energy. Then V in the 
driving term can be replaced by $G_0^{-1}$. Using that and the relation 
(\ref{e20}) we find

\begin{eqnarray}
\label{e23}
U &=& PG_0^{-1} + P\tilde T + V_4^{(1)}(1+P)\phi +  
V_4^{(1)}(1+P)G_0\tilde T
\end{eqnarray}
This is the equation which provides U once $\tilde T$ has been found. 

It remains to exhibit the formal connection to the previously used forms 
(\ref{e8})-(\ref{e10}). 

We use now the following property of $V_4$ underlying its decomposition used 
throughout:

\begin{eqnarray}
\label{e24}
(1+P)V_4 &=& (1+P)V_4^{(1)}(1+P)
\end{eqnarray}
Let us then define

\begin{eqnarray}
\label{e25}
\hat V_4 &\equiv& V_4^{(1)}(1+P)
\end{eqnarray}
and
\begin{eqnarray}
\label{e26}
\hat t_4 &=& \hat V_4 + \hat V_4 G_0 \hat t_4
\end{eqnarray}
Then it follows

\begin{eqnarray}
\label{e27}
(1+P)t_4 &\equiv& (1+P)[V_4 + V_4G_0V_4 +V_4G_0V_4G_0V_4 + \cdots] \nonumber \\
~ &=& (1+P)V_4^{(1)}(1+P) + (1+P)V_4^{(1)}(1+P)G_0V_4 + \cdots  \nonumber    \\
~ &=& (1+P)[V_4^{(1)}(1+P) + V_4^{(1)}(1+P)G_0V_4^{(1)}(1+P) +\cdots \nonumber \\
~ &=& (1+P)[ \hat V_4 + \hat V_4 G_0\hat V_4 + \cdots]  \nonumber \\
~ &\equiv& (1+P)\hat t_4
\end{eqnarray}
Further we introduce 

\begin{eqnarray}
\label{e28}
\tau_4 &\equiv& \hat t_4 + \hat t_4 G_0 T \nonumber \\
~ &=& \hat V_4 + \hat V_4 G_0 \tau_4 + \hat V_4 G_0 T
\end{eqnarray}
Applying $(1+P)$ from the left we get

\begin{eqnarray}
\label{e29}
(1+P)\tau_4 &=& (1+P)t_4 + (1+P)t_4G_0T \\
~           &=& T_4 \nonumber
\end{eqnarray}
according to (\ref{e8}).

Let us now add the information from the first equation of the set (\ref{e8}) 
to get

\begin{eqnarray}
\label{e30}
T + \tau_4 &=& tP + tG_0(1+P)\tau_4 + tPG_0T + \hat V_4 + \hat V_4G_0\tau_4 
+ \hat V_4G_0T
\end{eqnarray}

Using again (\ref{e28}) yields 

\begin{eqnarray}
\label{e31}
\nonumber
\tilde T &=& tP + \hat V_4 + tG_0P\tilde T + tG_0\hat V_4 + 
tG_0\hat V_4G_0\tilde T + \hat V_4G_0\tilde T \\
~ &=& tP + (1+tG_0)\hat V_4 + tG_0P\tilde T + (1+tG_0)\hat V_4G_0\tilde T
\end{eqnarray}
with
\begin{eqnarray}
\label{e32}
\tilde T &\equiv& \tau_4 + T
\end{eqnarray}
This Eq.(\ref{e31}) is identical to (\ref{e21}).

Now the full breakup operator as given in (\ref{e9}) is 

\begin{eqnarray}
\label{e33}
U_0 &=& (1 + P) T + T_4  \nonumber \\
~   &=& (1+P)T + (1+P)\tau_4 \nonumber \\
~   &=& (1+P)\tilde T
\end{eqnarray}
as required by (\ref{e19}).

Finally the operator for elastic scattering (\ref{e10})
can be rewritten, using (\ref{e29}) and (\ref{e28}) as 

\begin{eqnarray}
\label{e34}
U  &=& PG_0^{-1} + PT + T_4\nonumber \\
   &=& PG_0^{-1} + P\tilde T + \tau_4  \nonumber \\
~  &=& PG_0^{-1} + P\tilde T + \hat V_4 + \hat V_4 G_0 \tilde T
\end{eqnarray}
which is identical to (\ref{e23}). This concludes the verification of the 
equivalence of the old and new formulations. In practice we use the second
form of Eq.(\ref{e34}) with the quantity $\tau_{4}$ to evaluate $U$, since it
occurs naturally in Eq.(\ref{e21}) as an intermediate amplitude.

We developed a computer code for the new forms (\ref{e21}), (\ref{e19}) and 
(\ref{e23}). This has significant practical  advantages over the previous 
formulation. 
In the new approach one is simply iterating (\ref{e21}), typically 18 times 
for total angular momentum and parity $1/2^{+}$  
followed by a Pad\'e summation. The number of iterations for $1/2^{-}$ is
smaller and decreases further with increasing angular momentum.
In  the old scheme the set (\ref{e8}) expanded
in power of $V_4$ (see \cite{ref10}) had to be iterated. Since typically for
$1/2^{+}$ 10 powers of $V_4$ had to be kept and for each order of $V_{4}$ one
has to iterate the corresponding Faddeev equations typically 13 times, the
number of iterations in the new approach is drastically reduced. However, in
the new approach the kernel in Eq.(\ref{e21}) includes $V_{4}$ dependent parts
and requires therefore more CPU time for its evaluation. Altogether we found
in a realistic calculation a reduction of about a factor 4 in the CPU time for
the new approach.

We demonstrate now the feasibility of the new formalism and the numerical 
accuracy achieved with that form by showing a few 3N scattering observables 
evaluated with the old and new code. These are fullfledged calculations 
based on the NN potential CD Bonn (np) \cite{ref17} kept up to $j_{max}=2$ (j
is the total  
2-body angular momentum) and the Tucson-Melbourne $\pi-\pi$ exchange 
3NF\cite{ref4}, also kept up to $j_{max}=2$. 
The cut-off parameter in the 3NF has been chosen as $\Lambda=5.8\
m_{\pi}$. 

Fig.~\ref{fig1} shows two spin observables in elastic nd scattering
at 10.3 MeV. We have chosen cases where the effect of the 3NF is large. (The
elastic differential cross section shows no effect of the 3NF at all.) The two
curves with the 3NF included and evaluated in the old and new scheme overlap
perfectly. This is generally true for all elastic observables.

For the breakup process we show again two examples in Fig.~\ref{fig2} with
large 3NF effects. One is
a breakup cross section under np QFS conditions and the other a deuteron
analyzing power under the same conditions. For that very small observable
$A_{y}$ one can see tiny differences between the predictions of the old and
new scheme. This is the worst case we found by checking in the standard
breakup configurations cross sections and all analyzing powers, spin
correlation coefficients and vector spin transfer coefficients.

Summarizing, we have introduced a new single equation for 3N scattering 
including a 3NF. It is a direct generalisation of what has already 
been used for the 3N bound state. It is significantly faster to solve than 
our previous form (\ref{e8}), used up to now\cite{ref11}. We 
demonstrated the feasibility and accuracy of the new form.
 
\begin{acknowledgements}
This work was supported by the Deutsche Forschungsgemeinschaft (D.H.), the
Science  
and Technology Cooperation Germany-Poland under Grant No. XO81.91,  and 
by the Polish Committee for Scientific Research 
under Grant No. PB 1031. The numerical calculations have been performed 
on the Cray T90 of the H\"ochstleistungsrechenzentrum in J\"ulich, 
Germany.
\end{acknowledgements}

\newpage

\begin{figure} [h]
\caption
{\label{fig1} 
Fig. 1: Spin correlation coefficient $C_{yy}$ and neutron to deuteron
  spin transfer coefficient $K_{y}^{x'z'}$ at $E_N^{lab}=10.3$ MeV. The solid
  curve is without 3NF and the long and short dashed curves are with 3NF
  evaluated with the old and new approach, respectively.}
\end{figure}

\begin{figure}[h]
\caption
{\label{fig2} 
Fig. 2: The differential cross section and the deuteron analyzing power at
$E_N^{lab}=10.3$ MeV under np QFS conditions as a function of the scattering
angle of one nucleon. Curves as in Fig. 1}
\end{figure}
 
\end{document}